\documentclass[aps,prd,nofootinbib,longbibliography,10pt]{revtex4-2}
\usepackage{etoolbox}
\usepackage{amsmath,amssymb,amsfonts,mathtools}
\usepackage{mathrsfs,bbold}
\usepackage{graphicx}
\usepackage[colorlinks=true,urlcolor=blue,citecolor=red,linkcolor=blue]{hyperref}
\usepackage{accents}
\newlength{\dhatheight}
\usepackage{natbib}
\usepackage{wrapfig}
\usepackage{flushend,BOONDOX-cal,BOONDOX-frak}

\makeatletter
\newsavebox{\@brx}
\newcommand{\llangle}[1][]{\savebox{\@brx}{\(\m@th{#1\langle}\)}%
  \mathopen{\copy\@brx\kern-0.5\wd\@brx\usebox{\@brx}}}
\newcommand{\rrangle}[1][]{\savebox{\@brx}{\(\m@th{#1\rangle}\)}%
  \mathclose{\copy\@brx\kern-0.5\wd\@brx\usebox{\@brx}}}
\makeatother

\begin{document}
\title{{\bf{Probing the massive scalar mode in the levitated sensor detector of gravitational wave}}}
\author{Rakesh Das}
\email{rakeshkrishno@gmail.com}
\affiliation{Department of Physics, West Bengal State University, Barasat, Kolkata 700126, India}
\author{Anirban Saha}
\email{anirban@wbsu.ac.in}
\affiliation{Department of Physics, West Bengal State University, Barasat, Kolkata 700126, India}
\begin{abstract}
\noindent   
Owing to the mass scale associated with it the scalar longitudinal polarization mode of gravitational wave predicted in various modified theories of gravity should propagate at a subluminal speed 
and thus arrive with a time delay (for burst signals) or a phase difference (for persistent signals) at the detector site compared to the massless tensor polarization modes which move at the speed of light and are present in both standard general relativity and modified theories.  The longitudinal massive scalar mode interacts non-trivially with detectors along the signal propagation direction in contrast to massless the tensor modes which interact only in the transverse plane.  Identifying the signature of these distinctive features in a gravitational wave signal can provide observational evidence in favour of modified theories of gravity over general relativity. 
In this work we argue that owing to its compact design and tunability of operational frequency the recently proposed levitated sensor detectors \cite{Aggarwal} that works on the principle of gravitational wave induced resonant oscillation of a optically trapped \cite{Ashkin_1970} dielectric nanosphere sensor \cite{Geraci} can be useful in this regard.   We demonstrate that the dynamics of the levitated sensor mass obeys a geodesic deviation equation in the proper detector frame and construct a quantum mechanical description of this system in modified gravity framework to compute the probabilities of resonant transitions in response to incoming gravitational wave signals of both periodic and aperiodic kind.
\end{abstract}
\maketitle 
\section{Introduction}
Since the first gravitational wave (GW) was detected in 2015 \cite{1stGW_1,1stGW_2},  the detection events have become fairly regular \cite{GW_events_1,  GW_events_2} in the interferometric detectors like Advanced LIGO \cite{LIGO, pol5,  pol13,  pol17} and Advanced Virgo \cite{VIRGO}.  Apart form the area of GW astronomy \cite{GW_events_1,  GW_events_2,  advanceGW3,  gra_data1,  gra_data3,  gra_data4,  gra_data5,  gra_data6} these data can also be used to look into the innate nature of gravity itself,  especially since Einstein's general relativity (GR) is no more the only candidate theory of gravity at the classical level. The discovery of late time acceleration of the universe \cite{obe18,  obe19} has opened up possibilities for rival theories \cite{Alt_1,  Alt_2}.  A class of such alternatives go under the name of modified theory of gravity(MTG) \cite{MTG}.  

In a generic MTG framework matter couples to the gravity via the metric.  Linearising this theory yields GWs which can accommodate up to four extra polarisation modes in addition to the two tensor modes predicted by both MTG and standard GR \cite{Nishizawa09,  bogdanos,  clifford,  modes_2,  clifford2,  fr_re_1,  MTG_GW_massive,  fr_re_4}.  Among these additional ones,  the vector modes are usually ignored in theoretical analysis because only the scalar ones are allowed in most of the cosmologically viable scenarios \cite{Nishizawa09}.  

The mass scale associated with the scalar longitudinal mode ensures its propagation at a subluminal speed \cite{fr_re_1} and arrival with a time delay (for burst signals) or a phase difference (for persistent signals) at the detector site compared to the massless tensor modes which move at the speed of light.  Moreover the longitudinal massive scalar mode affects the detector along the signal propagation direction \cite{clifford,  modes_2,  clifford2} whereas the effects of the massless tensor modes are confined only in the transverse plane \cite{Nishizawa09,  bogdanos}.  This distinctive features of the longitudinal scalar polarization mode,  if identifid in a GW signal,  can provide observational evidence in favour of modified theories of gravity over GR.  

However to extract such polarisation content from a GW signal with sufficient precision  \cite{pol15,  pol16, Abbott_1,Abbott_2,Abbott_3,Takeda_1,Takeda_2,Hagihara} we need a full-grown network of widely seperated and simulteneously operating interferometric detectors.  Combining the data from multiple detector sites one can resolve the source direction and identify the transverse plane.  Moreover if several detector orientations could be arranged in a given site that would not only maximize the contribution of all possible polarization modes to the strain sensitivity but,  with the knowledge of propagation direction,  could also filter out the effect of longitudinal scalar mode,  if present in the signal.  But even if a network of interferometric detectors come online despite their operational budget,  having multiple detector orientation at a given site is not achievable due to their large-scale design. 

In comparsion the significantly compact design of the recently proposed levitated sensor detectors \cite{Aggarwal} of GW can be useful in this regard.  Here a dielectric nanosphere \cite{Geraci},  optically trapped \cite{Ashkin_1970} in the harmonic potential \cite{OT} at an antinode of a optical cavity,  serves as the sensor mass.  A passing GW displaces it from the minima of the potential.  The frequency of the  trapping potential can be varied by adjusting the intensity of the optical beam to achieve resonance\footnote{Although LSD depends on resonance mechanism,  the tunability of its operational frequency makes it suitable for scanning a wide range of the spectrum which the traditional resonant mass detectors \cite{bar_1,  bar_2, bar_3, bar_4,  bar_5} can not do. } with an incoming GW signal.  Currently under construction \cite{Aggarwal},  this detector claims to achieve improved sensitivity in a wider frequency range of $50-300$ kHz compared to Advanced LIGO \cite{LIGO,  pol5,  pol13,  pol17} while being relatively smaller in size $\sim 1-100\,\, {\rm m}$.  Naturally the underlying physics here is that of a quantum harmonic oscillator responding to GW\footnote{Since in any realistic scenario GW interacts with the test mass at a length scale of ${\cal{O}}\left[10^{-22} {\rm m}\right]$ a quantum mechanical treatment is of course preferable \cite{Weber_1,  Weber,  deWitt,  Kiefer,  eu}.  } .

Therefore in this paper we shall analyse the interaction of a quantum harmonic oscillator with GW in the MTG framework,  specifically focusing on the response of the test mass to the longitudinal scalar mode.  Since one of the kye idea is to have multiple LSDs arranged in different orientations of the optical cavity we shall consider a three-dimensional oscillator. This  is also necessary to incorporate the effect of the massive longitudinal scalar mode.  Note that not only the sensor mass but also the antinode position gets affected by the passing GW \cite{Geraci},  so a carefull analysis of this levitated sensor detector (LSD) geometry is necessary before we construct a quantum mechanical depiction of the detector-GW interaction in the proper detector frame.  So we shall first consider the basic schematics of the optical cavity and the sensor position of the LSD as described in \cite{Geraci} and establish that the time evolution of the displacement of the sensor from the antinode is indeed governed by a geodesic deviation equation in the rest-frame of the detector which serves as a Earth-bound proper detector frame.  Then employing a generic methodology we have elaborated earlier in \cite{Rakesh_1} we shall obtain a quantum mechanical description of the system to estimate the probabilities of the sensor undergoing GW-induced resonant transitions to some excited states \cite{maggiore,  NCGW_HO1,  NCGWSwarup,  NCGW_T_1,  NCGW_T_2,  NCGW_T_3}.  The GW signal will be assumed to contain both the massive scalar and massless tensor polarization modes predicted by the classical MTG framework.

This paper is structured as follows: In the next section we consider the basic geometry of the levitated sensor detector to establish that its response to the incoming GW can be tracked by studying a geodesic deviation equation.  In section 3 we switch to a quantum mechanical description of the system. The formal dynamical solution is provided in Appendix A.  In section 4 we obtain the transition probabilities to demonstrate the resonant transitions for some simple GW waveform templets.  We discuss the scope of levitated sensor detectors in view of our results in section 5.

\section{Geometry of LSD and the force of GW on the sensor}
In this section we consider the geometry of the optical cavity and the sensor trapped inside  following \cite{Geraci},  but in a manner suitable for our analysis\footnote{For the need of easy comparison with \cite{Geraci} we keep the notations as same as  possible. }.  The origin of the locally Lorentz frame (LLF),  i.e.  the TT gauge frame \cite{maggiore} is at the input mirror ${\rm M}_{1}$.  The spatial coordinates of the other mirror ${\rm M}_{2}$ located at the far end of the optical cavity are $x^{i}_{{\rm M}_2} = \left(l_{\rm m}, 0,0\right)$.  Thus the proper distance between the two mirrors in presence of GW is given by 
\begin{equation}
L_{\rm m} \left(t\right)= \left\{\left(\eta_{ij} + h^{\rm TT}_{ij}\right) \Delta x^{i} \Delta x^{j} \right\}^{\frac{1}{2}} \simeq \left(1 + \frac{1}{2} h_{11} \left( t \right)\right) l_{\rm m} + {\cal O}\left[h^{2}\right].
\label{Lm}
\end{equation}
where $t$ is the coordinate time of the LLF and $h_{11} \left( t \right)$ is the value of the relevant component of the metric perturbation $h^{\rm TT}_{ij}$ in TT gauge,  that contributes to changing the proper distances along the $x$-direction.

Similarly,  if the spatial coordinates of the sensor S are $x^{i}_{\rm S} = \left(x_{\rm S}, 0,0\right)$,  its proper distance from the origin during the passage of GW is 
\begin{equation}
X_{\rm S}\left(t\right)  \simeq \left(1 + \frac{1}{2} h_{11} \left( t \right)\right) x_{\rm S} + {\cal O}\left[h^{2}\right].
\label{Xs}
\end{equation}
Note that in absence of GW,  let's say initially,  at time $t = 0,$ both the proper distances are just equal to the corresponding coordinate readings with respect to the origin and coordinate readings do not change with time in TT gauge frame.
In the optically levitated sensor,  the dielectric nanosphere is trapped in the minima of a harmonic potential \cite{Geraci} located at one of the antinodes created by the interference of the two counter-propagating waves in the cavity.  In absence of GW,  (at time $t = 0$) the antinode position $x_{\rm min}$ is determined by the interference condition 
\begin{equation}
k_{t} \left(l_{\rm m} - x_{\rm min}\right) = \left(n +\frac{1}{2}\right) \pi
\label{an}
\end{equation}
where $n$ is an integer.
Since the antinode position $x_{\rm min}$ depends on the proper distance of the end-mirror ${\rm M}_{2}$,  at $t >0$,  it will change to $x^{\prime}_{\rm min}$ satisfying the interference condition in presence of GW,
\begin{equation}
k_{t} \left(L_{\rm m} - x^{\prime}_{\rm min}\right) = \left(n +\frac{1}{2}\right) \pi
\label{AN}
\end{equation}
So when GW passes,  the antinode shifts position by 
\begin{equation}
 \delta X_{\rm min} = x^{\prime}_{\rm min} -  x_{\rm min}  = \left(L_{\rm m} - l_{\rm m} \right)  = \frac{1}{2} h_{11} \left( t \right) l_{\rm m}
\label{Xmin}
\end{equation}
where the last equality follows from eqn. (\ref{Lm}).
The corresponding change in the position of the sensor is
\begin{equation}
 \delta X_{\rm S} = X_{\rm S} -  x_{\rm S} = \frac{1}{2} h_{11} \left( t \right) x_{\rm S}
\label{XS}
\end{equation}
 according to eqn.(\ref{Xs}).  So GW moves the sensor away from the aninode by 
\begin{equation}
 \Delta X = \delta X_{\rm S} -  \delta X_{\rm min} =  \frac{1}{2} h_{11} \left( t \right) \left( x_{\rm S} - l_{\rm m} \right)
\label{DX}
\end{equation}
and the force by GW on the sensor mass $m_{\rm S}$ can be cast in the Newtonian form as 
\begin{equation}
 F_{\rm GW} = m_{\rm S} \frac{d^{2}}{dt^{2}} \left( \Delta X \right) =  \frac{m_{\rm S}}{2} \ddot{h}_{11} \left( t \right) \left( x_{\rm S} - l_{\rm m} \right)
\label{GW_force}
\end{equation}
where dot denotes time derivative with restpect to the coordinate time of the LLF.   Neither $x_{\rm S}$ nor $l_{\rm m}$ changes with time since they are just coordinate readings in the LLF.

Now in the LLF both the sensor S and the mirror  ${\rm M}_{2}$ follow their respective geodesics $x^{i}_{\rm S}\left(\tau_{\rm S}\right)$ and $x^{i}_{{\rm M}_{2}}\left(\tau_{{\rm M}_{2}}\right)$ parametrized by their proper times $\tau_{\rm S}$ and $\tau_{{\rm M}_{2}}$ which are both identical to the coordinate time in TT gauge \cite{maggiore}. 

We take their seperation as the geodesic deviation $\xi ^{i}\left(t\right) = x^{i}_{\rm S} - x^{i}_{{\rm M}_{2}}$ which varies from its initial value $\xi ^{1}\left(0\right) = \left(x_{\rm S} - l_{\rm m} \right)$ at $t = 0$, when there is no GW,  to a later value $\xi ^{1}\left(t\right) = \left(X_{\rm S} - L_{\rm m} \right)$ at $t = t$,  in presence of GW. 

Using eqn.s (\ref{Xmin},  \ref{XS}) we can write the sensor to antinode seperation $\Delta X$,  defined in eqn. (\ref{DX}),  in terms of these geodesic deviations as
\begin{eqnarray}
 \Delta X = \left( X_{\rm S} -  L_{\rm m} \right) - \left(x_{\rm S} - l_{\rm m} \right)  = \xi ^{1}\left(t\right) - \xi ^{1}\left(0\right) = \frac{1}{2} h_{11} \left( t \right)\xi ^{1}\left(0\right)
\label{DX_1}
\end{eqnarray}
Note that at $t>0$,  the geodesic deviation $\xi ^{1}\left(t\right)$ is the difference of two proper distances $X_{\rm S}$ and $L_{\rm m} $,  so it varies with time.  Its initial value $\xi ^{1}\left(0\right) $ remains constant as it is just the difference of two coordinate values $x_{\rm S}$ and $l_{\rm m} $,  in the LLF,  as discussed below eqn.(\ref{Xs}).  Therefore we have 
\begin{eqnarray}
\frac{d^{2}}{dt^{2}} \left( \Delta X \right) = \ddot{ \xi}^{1}  \left(t\right)
\label{DX_00}
\end{eqnarray}
which shows that the dynamics of the shift $ \left( \Delta X \right)$ is mimicked by the geodesic deviation $\xi ^{1}\left(t\right)$.    

The rest frame of our optically levitated sensor is a proper detector frame where the geodesic deviation equation of the sensor mass $m_{\rm S}$,  trapped in a harmonic potential of intrinsic freequency $\omega$,  becomes 
\begin{equation}
m_{\rm S} \ddot{\xi}^{1} \left(t\right)=  -m_{\rm S} R^{1}_{0, 1 0} \xi^{1}\left(t\right) -  m_{\rm S} \omega^{2} \xi^{1}\left(t\right)
\label{GD}
\end{equation}
in the long wave-length and low velocity limit\footnote{The long wavelength ($|x| << \frac{\lambda}{2\pi}$,  the reduced wavelength of GW) and low-velocity (velocities of the test mass are non-relativistic $v << c$) limit is where all the earthbound GW detectors operate. }.  Since in a linearized theory,  the Reimann curvature tensor is an invariant \cite{maggiore},   its components evaluated in the TT gauge frame can be used in this proper detector frame.  When the relevant components of the curvature tensor\footnote{The coordinate time of the proper detector frame,  i.e.,  the proper time of the $\tau_{\rm S}$ of the sensor S is identical to the coordinate time $t$ of the LLF  in the TT gauge.}
$R^{1}_{0,  1 0} = -\frac{1}{2}\ddot{h}_{11}\left(t\right)$,  
are substituted in eqn.(\ref{GD}),  it takes the form of the equation of motion of a forced harmonic oscillator
\begin{equation}
m_{\rm S} \ddot{\xi}^{1}\left(t\right) =  \frac{m_{\rm S}}{2}\ddot{h}_{11}\left(t\right)  \xi^{1}\left(t\right) -  m_{\rm S} \omega^{2} \xi^{1}\left(t\right)
\label{GD_1}
\end{equation}
In (\ref{GD_1}) the force of GW in terms of the geodesic deviation $\xi^{1}\left(t\right)$ is identical to the $F_{\rm GW}$ on the sensor S we calculated earlier in eqn.(\ref{GW_force}) 
to ${\cal O}\left[h\right]$,  as 
\begin{equation}
\frac{m_{\rm S}}{2} \ddot{h}_{11} \left( t \right) \left( x_{\rm S} - l_{\rm m} \right) = \frac{m_{\rm S}}{2} \ddot{h}_{11} \left( t \right) \xi ^{1}\left(0\right) \simeq \frac{m_{\rm S}}{2} \ddot{h}_{11} \left( t \right) \xi ^{1}\left(t\right)   + {\cal O}\left[h^{2}\right]
\label{GW_force_1}
\end{equation}
This establishes that to 1st order in the $h_{11} \left(t\right)$,  the dynamics of the sensor-antinode seperation $\Delta X$ is indeed governed by the geodesic deviation equation (\ref{GD_1}).  Its three dimensional generalization is 
\begin{eqnarray}
m_{\rm S}\ddot{\xi}^{i}  =  \frac{m_{\rm S}}{2} \ddot{h}_{ij} \left( t \right)  \xi^{j} - m_{\rm S} \omega^{2} \xi^{i}
\label{DX_00_i}
\end{eqnarray}
which we use in the next section to obtain a quantum mechanical description of the system. 
\section{Quantum mechanics of the LSD-GW Interaction }
The  Lagrangian that leads to the classical equation of motion (\ref{DX_00_i}) of the sensor mass $m_{\rm S}$ is 
\begin{eqnarray}\label{lagrangian}
\mathcal{L}_{\rm S}=\frac{1}{2}m_{\rm S} \left(\dot{\xi^{i}}\right)^{2} - m_{\rm S} \Gamma^{i}_{0j} \dot{\xi}_{i} \xi^{j} - \frac{1}{2} m_{\rm S} \omega^{2} \left(\xi^{i}\right)^{2}
\end{eqnarray}
up to a total time-derivative term.  Owing to its interaction with the GW,  the sensor mass picks up contribution from the affine connections $\Gamma^{i}_{0j} =\frac{1}{2}\dot{h}_{ij}$ in its canonical momentum $p_{i}= \left( m_{\rm S}\dot{\xi}_{i} - m_{\rm S}\Gamma^{i}_{0j}\xi^{j}\right)$.  Thus its Hamiltonian,  to first order in $h_{ij}$,  is given by 
 \begin{eqnarray}\label{hamilton_1}
\mathcal{H}_{\rm S}=\sum _{i}\left[ \frac{p_{i}^{2}}{2m_{\rm S}} + \frac{1}{2} m_{\rm S} \omega^{2} \xi_{i}^{2} + \sum_{j}\frac{\Gamma^{i}_{0j}}{2} \left(\xi_{i} p_{j} + p_{i} \xi_{j}\right)\right]
\end{eqnarray}
where $\xi_{i}$ and $ p_{i}$ are the pair of canonical variables for the sensor.  By replacing them with the corresponding position and momentum operators $\left(\hat{\xi}_{i},  \hat{p}_{i}\right)$ which satisfy the Heisenberg algebra,  we can quantize the sensor Hamiltonian and study its quantum mechanical behaviour.  In terms of the  raising and lowering operators $a_{i}$ and $a_{i}^{\dagger}$,  defined as
\begin{align} \label{operator}
\hat{\xi}_{i}=\left(\frac{\hbar}{2m_{\rm S}\omega}\right)^\frac{1}{2}\left(a_{i}+a_{i}^{\dagger}\right);~~
\hat{p}_{i}=-i\left(\frac{m_{\rm S}\omega\hbar }{2}\right)^{\frac{1}{2}}\left(a_{i}-a_{i}^{\dagger}\right)
\end{align}
the quantum mechanical Hamiltonian for the sensor assumes the form 
\begin{eqnarray}\label{hamilton_HO}
\hat{\mathcal{H}}_{\rm S}=\sum_{i}\hbar \omega (a_{i}^\dagger a_{i}+\frac{3}{2})-\sum_{i,  j}\frac{i\hbar}{4}\dot h_{ij}(t)\left(a_{j}a_{i} - a^{\dagger}_{j}a^{\dagger}_{i}\right)
\end{eqnarray}
which dictates the time evolution through the Heisenberg equations of the operators $a_{i}(t)$ and $a^{\dagger}_{i}(t)$ given by 
\begin{eqnarray}\label{operator3}
\frac{da_{i}(t)}{dt} = \frac{i}{\hbar}\left[\hat{\mathcal{H}}, a_{i}\right] 
\end{eqnarray}
and its hermitian conjugate respectively.  In Appendix \ref{AppA} we solve for $a_{i}(t)$ and $a_{i}^{\dagger}(t)$ in terms of their initial values $a_{i}(0)$ and $a_{i}^{\dagger}(0)$ and express the formal solution as expectation values of the position components of the sensor $\langle{\xi}_{i}(t)\rangle$.  In the reminder of the paper we use time dependent perturbation theory to compute the probability of GW induced resonant transitions of the sensor mass.
\section{ Probabilities for GW-induced transitions}
The Hamiltonian $\mathcal{H}_{\rm S}$ (\ref{hamilton_HO}) is a sum of the unperturbed Hamiltonian of a quantum harmonic oscillator
\begin{eqnarray}
 \mathcal{H}_0 &=& \sum_{j}\hbar \omega (a_j^\dagger a_j+\frac{3}{2})
 \label{H_0}
 \end{eqnarray}
 which represents the sensor mass trapped in the three-dimensional harmonic potential of the optical cavity and a perturbation term
\begin{eqnarray}\mathcal{H}_{in} &=&  \sum_{j,k} \frac{i\hbar}{4}\dot h_{jk}(t)(a^{\dagger}_{j}a^{\dagger}_{k}-a_{j}a_{k})
    \label{H_int}
\end{eqnarray}
 describing its interaction with a linearly polarized GW.  The unperturbed energy eigenstates and eigenvalues for the ground state and the first few excited states of the oscillator Hamiltonian $\mathcal{H}_0$ are 
\begin{table}[h!]
\centering
\renewcommand{\arraystretch}{1.5} 
\begin{tabular}{|c|c|}
\hline
Eigenstates & Eigenvalues \\
\hline
$\Psi_{0} = | 0,0,0\rangle$ & $\frac{3}{2}\omega\hbar$ \\
\hline
$\Psi_{1} = \frac{1}{\sqrt{3}} \left( | 0,0,1\rangle + | 0,1,0\rangle + | 1,0,0\rangle \right)$ & $\frac{5}{2}\omega\hbar$ \\
\hline
$\Psi_{2} = \frac{1}{\sqrt{6}} \left( |2, 0, 0\rangle + |0, 2, 0\rangle + |0, 0, 2\rangle + |1, 1, 0\rangle + |1, 0, 1\rangle + |0, 1, 1\rangle \right)$ & $\frac{7}{2}\omega\hbar$ \\
\hline
\begin{tabular}{@{}c@{}} 
$\Psi_{3} = \frac{1}{\sqrt{10}} \left( |3, 0, 0\rangle + |0, 3, 0\rangle + |0, 0, 3\rangle + |1, 2, 0\rangle + |2, 1, 0\rangle \right.$ \\ 
$\left. + |0, 2, 1\rangle + |0, 1, 2\rangle + |2, 0, 1\rangle + |1, 0, 2\rangle + |1, 1, 1\rangle \right)$ 
\end{tabular} & $\frac{9}{2}\omega\hbar$ \\
\hline
\end{tabular}
\caption{Unperturbed Eigenstates and Corresponding Energy Eigenvalues}
\label{tab:example}
\end{table}
\newpage
\noindent The amplitude of transition of the sensor from an initial state $|i>$ to a final state $|f>$, ($i \neq f $) to the lowest order approximation is  
\begin{eqnarray}
    C_{i\rightarrow f}(t\rightarrow \infty) = -\frac{i}{\hbar}\int^{\infty}_{-\infty} dt'e^{\frac{i}{\hbar}(E_{f}-E_{i})t'}\sum_{j,k}F_{jk}(t')<\Psi_{f}|\hat{Q}_{jk}|\Psi_{i}>
   \label{tran_pro}
\end{eqnarray}
where time-dependent perturbing Hamiltonian $\mathcal{H}_{in} = F_{jk}(t')\hat{Q}_{jk}$ is factored into $F_{jk}(t') = \frac{i\hbar}{4}\dot h_{jk}(t')$ that contains the explicit time dependence and $\hat{Q}_{jk} = ( a^{\dagger}_{j}a^{\dagger}_{k} -  a_{j}a_{k} )$ involving the ladder operators that act on the eigenstates.

Using equation (\ref{tran_pro}) the first two most prominant transitions turn out to be $\Psi_{0} \rightarrow \Psi_{2}$ and  $\Psi_{1}\rightarrow\Psi_{3}$.  The transition amplitudes are
\begin{eqnarray}
C_{0\rightarrow 2}(t \rightarrow \infty) =  \int^{\infty}_{-\infty}  dt' e^{2i\omega t'}\left[\frac{1}{\sqrt{48}}\left(\dot{h}_{11}+\dot{h}_{22}+\dot{h}_{33}\right) + \frac{1}{\sqrt{24}}\left(\dot{h}_{12}  +\dot{h}_{13}+\dot{h}_{23}\right)\right]
\label{c02_f}
\end{eqnarray}
and 
\begin{eqnarray}
C_{1\rightarrow 3}(t \rightarrow \infty) = \int^{\infty}_{-\infty} dt' e^{2i\omega t'}\left[\frac{2+\sqrt{3}}{2\sqrt{60}}\left(\dot{h}_{11}+\dot{h}_{22}+\dot{h}_{33}\right) + \frac{2\sqrt{2}+1}{2\sqrt{30}}\left(\dot{h}_{12}  +\dot{h}_{13}+\dot{h}_{23}\right)\right]
\label{c13_f} 
\end{eqnarray}
These amplitudes and the corresponding transition probabilities $P_{i \rightarrow f} = |C_{i \rightarrow f}|^{2}$ depend on the waveforms $h_{jk}\left(t \right)$ of the GW signal that perturbs the sensor mass away from its equilibrium position at the anti-node of the optical cavity.  When the frequency of the GW signal and the intrinsic frequency $\omega$ of the trapping potential meets the resonance condition the sensor mass will undertake  resonant transitions.  To demonstrate this explicitly we take some sample GW waveform templates of both periodic and aperiodic kind that are likely to be generated in runaway astronomical events \cite{maggiore}.  Since our main purpose is to anticipate the signature of scalar mode of GW predicted by MTG in these transitions,  the massive longitudinal scalar mode is of course incorporated in the waveform along with the usual massless tensorial modes of polarization.  
\subsection{Periodic linearly polarized GW:}
To express the typical form of the linearly polarized periodic GW signal originating from a specific astronomical sourse in a MTG framework we choose the $z-$axis of our proper detector frame along the direction of the signal propagation,  so that the polarization characteristics of the signal are contained in the three independent polarization tensors~\cite{Nishizawa09,clifford,  modes_2,  clifford2}
\begin{eqnarray} \label{pol_basis}
\wp^{1}_{jk} = \wp^{\times}_{jk}=
\begin{pmatrix}
0 & 1 & 0 \\
1 & 0 & 0 \\
0 & 0 & 0
\end{pmatrix}, ~~~~~~
\wp^{2}_{jk}=\wp^{+}_{jk}=
\begin{pmatrix}
1 & ~0 & 0 \\
0 & -1 & 0 \\
0 & ~0 & 0
\end{pmatrix},~~ 
\wp^{3}_{jk}=\wp^{{\rm s}}_{jk}=\sqrt{2}
\begin{pmatrix}
0 & 0 & 0 \\
0 & 0 & 0 \\
0 & 0 & 1
\end{pmatrix}.
\end{eqnarray}
The tensor modes vary sinusoidally with a frequencies $\Omega$ and propagate with the speed of light along the wave vector $\vec{k}= k\hat{z}$,  such that $\Omega = |\vec{k} | = k$.   The massive scalar mode either has
the same propagation vector $\vec{k} = k\hat{z}$ but a different frequency $\Omega_{\rm s}$,  owing to the massive disperssion relation $ 2 \pi \Omega_{s} = \sqrt{c^{2} k^{2} + \left(\frac{mc^{2}}{\hbar}\right)^{2}}$,  so that the GW signal takes the form 
\begin{equation}\label{GW_periodic}
h_{jk} = \left(A_{+}(\vec{k})\wp^{+}_{jk} + A_{\times}(\vec{k})\wp^{\times}_{jk}\right) \cos\Omega t + A_{\rm s}(\vec{k})\,\wp^{\rm s}_{jk} \cos\Omega_{\rm s} t
\end{equation}
 or, 
the same frequency $\Omega$ but a different propagation vector $\vec{k}_{\rm s}= k_{\rm s}\hat{z}$ in the same direction but with a different magnitude,  satisfying the disperssion relation $ 2 \pi \Omega = \sqrt{c^{2} k_{\rm s}^{2} + \left(\frac{mc^{2}}{\hbar}\right)^{2}}$,  leading to the GW signal 
\begin{equation}\label{GW_periodic_same}
h_{jk} = \left(A_{+}(\vec{k})\wp^{+}_{jk} + A_{\times}(\vec{k})\wp^{\times}_{jk}\right) \cos\Omega t + A_{\rm s}(\vec{k}_{\rm s})\,\wp^{\rm s}_{jk} \cos\Omega t
\end{equation}
Note that in both the scenarios the subluminal propagation speed of the massive scalar mode 
\begin{eqnarray} 
c_{\rm s}  &=& \begin{cases}   \frac{\partial }{\partial k} \left(2 \pi\Omega_{\rm s}\right)= c \left[1 - \left(\frac{mc^{2}}{h \Omega_{\rm s}}\right)^{2}\right]^{\frac{1}{2}}\,\, \left[{\rm different  \,\, frequency}\right]\nonumber \\ 
 \frac{\partial }{\partial k_{\rm s}} \left(2 \pi\Omega \right)= c \left[1 - \left(\frac{mc^{2}}{h \Omega} \right)^{2}\right]^{\frac{1}{2}}\,\, \left[{\rm different  \,\, wave \,vector\,\, magnitude}\right]
\end{cases} \\ 
    \label{c_s}
\end{eqnarray}
is determined by its frequency $\Omega_{\rm s}$ or $\Omega$ and mass scale $m$ which characterizes how modified a specific MTG is from the standard GR.  

Now given a periodic emission process at the astronomical source,  we expect both the tensor and the scalar modes of the GW signal to be periodic with the same frequency and to compensate for its massive disperssion relation the wave vector of the scalar mode must have a different magnitude \cite{amin}.  In that case only the phase of the scalar mode,  owing to its subluminal speed,  will differ from that of the tensor modes in case of periodic signals.  However till date the currently operating GW detectors have only detected the transient,  short-duration signals from black-hole or nutron star mearger phenomena.  These are typically impulsive signals for which the scalar mode is expected to differ from tensor modes in arrival time.  On the other hand there are also several examples in the literature \cite{clifford2,  Maggiore2000,  DeFelice2010,  Capozziello2011} where the assumption of equal propagation vectors but different frequencies for the massive scalar and tensor modes of GW have been made.  Since this provides the additional scope to differenciating the massive scalar modes from the tensor ones in the resonant transition data,  in the reminder of this work we shall stick to the ``different frequency'' scenario and work with (\ref{GW_periodic}) as the GW signal.  Since all the transition probabilities are expressed in terms of the frequencies in this work,  the results for the ``same frequency'' scenario can be easily obtained by simply using $\Omega = \Omega_{\rm s}$.  Also since the arrival time of the signals for the scalar and tensor modes differ even for the ``same frequency'' scenario,  it is useful to keep track of the corresponding transition probabilities seperately.

Utilizing the equations (\ref{c02_f},\ref{c13_f}),  for the periodic perturbation (\ref{GW_periodic}) we get the amplitudes for the first two non-zero  transitions to be
\begin{eqnarray}
 C_{0\rightarrow 2} &=&    \frac{i}{2 \sqrt{24}}\Omega A_\times \left\{ \delta(2\omega +\Omega) - \delta(2\omega -\Omega) \right\} +   \frac{i}{2 \sqrt{24}} \Omega_{s} A_s \left\{ \delta(2\omega +\Omega_{s}) - \delta(2\omega -\Omega_{s}) \right\} 
 \label{tran_ampli_02}\\
C_{1\rightarrow 3} &=&   i \left(\frac{2\sqrt{2}+1}{4\sqrt{30}} \right) \Omega A_\times \left\{ \delta(2\omega +\Omega) - \delta(2\omega -\Omega) \right\} +  i\left( \frac{2+\sqrt{3}}{4\sqrt{30}} \right) \Omega_{s}A_s \left\{ \delta(2\omega +\Omega_{s}) - \delta(2\omega -\Omega_{s}) \right\}  \nonumber\\ 
\label{tran_ampli_13}
\end{eqnarray}
and the corresponding transition probabilities are
\begin{eqnarray}
 P_{0\rightarrow 2} &=&  \left[ \frac{1}{2 \sqrt{24}}\Omega A_\times \left\{ \delta(2\omega +\Omega) - \delta(2\omega -\Omega) \right\} +  \frac{1}{2 \sqrt{24}} \Omega_{s} A_s \left\{ \delta(2\omega +\Omega_{s}) - \delta(2\omega -\Omega_{s}) \right\}  \right]^2 
 \label{tran_probab_02}\\
 P_{1\rightarrow 3} &=&   \left[ \left(\frac{2\sqrt{2}+1}{4\sqrt{30}} \right)  \Omega A_\times \left\{ \delta(2\omega +\Omega) - \delta(2\omega -\Omega) \right\} +  \left( \frac{2+\sqrt{3}}{4\sqrt{30}} \right) \Omega_{s} A_s \left\{ \delta(2\omega +\Omega_{s}) - \delta(2\omega -\Omega_{s}) \right\}  \right]^2 \nonumber\\
 \label{tran_probab_13}
\end{eqnarray}
Owing to the physical constraint that all frequencies involved must be positive,  from (\ref{tran_probab_02},  \ref{tran_probab_13}) we can easily read off the resonance conditions for transition induced by the tensor modes $\left(2\omega = \Omega \right) $ and that by the scalar mode $\left( 2\omega = \Omega_{s}\right) $ of GW.  If the frequencies of the massless tensor mode $\Omega $ and massive scalar mode $\Omega_{s}$ are different\footnote{If the tensor and scalar modes arrive with the same frequency and propagation direction,  but with two seperate propagation vector magnitudes then there will be a singular resonance point at $2 \omega = \Omega = \Omega_{\rm s} $ where transition between a given pair of states triggered by the tensor mode will be seperated from that triggered by the scalar mode only by their phase difference. } for GW signal coming from a particular source with the same propagation vector $\vec{k}$,  they will show up as two distinct resonant transitions with a non-zero arrival time gap with the probabilities 
\begin{eqnarray}
\lim_{T \rightarrow \infty} \frac{1}{T} ~~   P_{0\rightarrow 2} &=& \begin{cases} \frac{1}{96} \Omega^2 A^2_{\times} \, \delta(2\omega-\Omega) ,  \left[{\rm tensor \,\, mode}\right]\nonumber \\ 
     \frac{1}{96}\Omega_{\rm s}^2 A^2_{s} \,\,\delta(2\omega-\Omega_{\rm s})  ,  \left[{\rm scalar \,\, mode}\right]  
    \end{cases} \\ 
    \label{TP_02}
\end{eqnarray}
for the transition from the ground state to the second excited state and
\begin{eqnarray}
\lim_{T \rightarrow \infty} \frac{1}{T} ~~   P_{1\rightarrow 3} &=& \begin{cases} \left( \frac{9+4\sqrt{2}}{480}\right)\Omega^2 A^2_{\times} \, \delta(2\omega-\Omega)  ,  \left[{\rm tensor \,\, mode}\right]\nonumber \\
   \left(\frac{7+4\sqrt{3}}{480}\right) \Omega_{\rm s}^2 A^2_{\rm s}\, \,\delta(2\omega-\Omega_{\rm s}) ,  \left[{\rm scalar \,\, mode}\right] 
    \end{cases} \\
 \label{TP_13}
    \end{eqnarray}   
for the transition from the first to the second excited state.  Here we have regularized $ \delta(0)$,  coming from the square of the delta function $\left[\delta(2\omega -\Omega)\right]^{2} = \delta(2\omega -\Omega)\delta(0),$ as $\delta\left(0\right) = T $ for a finite observation time $ -\frac{T}{2}<t<\frac{T}{2}$ using 
\[
\delta(\omega) =  \int^{\frac{T}{2}}_{-\frac{T}{2}} dt\, e^{i\omega t} 
\]
since in any realistic observational setup GW signals are typically observed over a finite time interval only.  

These expressions encapsulating the rates of specific quantum transitions show that the probability of transitions increase with the frequency of the GW signal quadratically.  Since the efficiency of the levitated sensor detector improves at the higher side of frequencies \cite{Aggarwal} it should be suitable to identify such transitions. 

Also note that  in the LSD \cite{Geraci,  Aggarwal} the frequency $\omega$ of the harmonic trapping potential can be tuned across a wide range of $50-300$ kHz by verying the intensity of the leser beams in the optical cavity.  In contrast the traditional resonant mass detectors \cite{bar_1,  bar_2, bar_3, bar_4,  bar_5} can operate only at their fixed natural resonant frequency decided by the length of the bar and speed of sound in the bar material.  But for a resonant response the detector has to be tuned to a ``right frequency at the right time'' which can only be achieved by a proper scanning stratagy,  especially for persistant periodic signals.  Naturally this makes LSD batter suited to aim for resonant detections,  although their still remain caveats like selecting the scanning speed,  frequency bandwidth to be covered in a year with sufficient sensitivity and most importantly,  identifying a stratagy to target the transient signals,  which are way more likely to arrive with sufficient signal strength to be detected over the noise floor.  

Given that a scanning stratagy is in any case essential,  if the tensor and scalar modes of a persistant GW signal arrive with different frequencies both the resonances may get captured in a set up with multiple LSDs. Since for a given transition it rates induced by the tensor and scalar modes depend on their frequencies and amplitudes if both the resonances are indeed detected-- $\Omega$ and $\Omega_{s}$ will be known,  so the relative strength of the tensor and scalar modes can be obtained by comparing the corresponding transition rates.  This will help model the source generating the tensor and scalar modes of GW.  The mass scale $m$ associated with the scalar mode that characterizes how modified a given MTG is from the standard GR,  can also be estimated from the resonance data using the relation $\left(\frac{m c^{2}}{h}\right)^{2}= \Omega_{s}^{2} - \Omega^{2}$.  Independently the difference of arrival time (phase difference) of the tensor and scalar modes \cite{Inagaki} can be used to constraine this mass scale which can then be corroborated with its estimation from the resonances data.
More challenging will be the scenario where the tensor and massive scalar modes have the same frequency,  for detecting the difference in arrival time for transient signals and phase difference in persistent periodic signals will be crucial to seperate the two modes.

\subsection{Aperiodic linearly polarized GW:}
\noindent  Aperiodic GW signals are expected from supernova explosions or inspiraling neutron star (or black hole) binaries at their last stable orbit or during their merger and final ringdown.  They emit GW in a burst with a huge amount of energy within a very short duration $10^{-4} {\rm sec}-1 {\rm ~ sec}$.  Such burst signals can be approximately modeled as\footnote{For reasons mentioned earlier in the section also we keep the propagation vector of the tensor and scalar modes identical and their frequencies distinct.}
\begin{eqnarray}
h_{jk} \left(t\right) =  g \left( t \right) \left(A_{\times}(\vec{k})\wp^\times_{jk} + A_{+}(\vec{k})\wp^+_{jk}\right) +  s \left( t \right)   A_{s}(\vec{k})\,\wp^s_{jk}
\label{lin_pol_burst}
\end{eqnarray}
where the time dependence of the tensor and scalar mode amplitudes $g\left( t \right)$ and $ s\left( t \right)$ are choosen to be Gaussian functions
\begin{eqnarray}
g \left(t\right) &=& f_{g}e^{- t^{2}/ \tau_{g}^{2}}\nonumber\\
s \left(t\right) &=& f_{s}e^{- \left(t+\Delta t\right)^{2}/ \tau_{s}^{2}}
\label{burst_waveform_Gaussian}
\end{eqnarray} 
that go to zero smoothly but rather fast beyond their respective temporal durations $|t| > \tau_{ g}$ and $ |t| > \tau_{s} $. 
Because these durations are small burst signals for both the tensor and scalar modes contain a wide range of frequencies\footnote{This should increase the chance of being picked up by the scanning stratagy to be employed in LSDs.} \cite{maggiore}.  From the maximum spread $\nu_{\rm max}$ of the broad range of continuum spectrum $\tau_g$ and $\tau_s$ can be estimated using $ \tau \sim \frac{1}{\nu_{\rm max}}$.  To ensure that GW signal hits the resonant sensor at $t=0$,  we  can put the additional condition $f_{g}|_{t = 0} = f_{s}|_{t = 0} =0$.  $\Delta t$ in(\ref{burst_waveform_Gaussian}) signifies the delay in the arrival time of the massive scalar mode due to its subluminal speed compared to the massless tensor mode.
The GW bursts (\ref{lin_pol_burst}) can be decomposed into their Fourier components as  
\begin{eqnarray}
h_{jk} \left(t\right) = \frac{1}{2 \pi} \left(A_{\times}\wp^\times_{jk} + A_{+}\wp^+_{jk}\right)  \int_{-\infty}^{+\infty} \tilde{g} \left( \Omega \right) e^{- i \Omega t}  d \Omega  + \frac{1}{2 \pi} A_{s}\wp^s_{jk}  \int_{-\infty}^{+\infty} \tilde{s} \left( \Omega_{s} \right) e^{- i \Omega_{s} t}  d \Omega_{s} 
\label{lin_pol_burst_Gaussian}
\end{eqnarray}
where the components themselves,  for the tensor and scalar modes at frequencies $\Omega$ and $\Omega_{s}$,  are respectively\footnote{Though in the Fouries expansion these frequencies are just dummy variables,  we should distinguish them since the burst signals for the massless tensor and massive scalar modes have different dispersion relations.} 
\begin{eqnarray}
\tilde{g} \left( \Omega \right) &=& \sqrt{\pi} f_{g}\tau_{g} e^{- \left( \frac{\Omega \tau_{g}}{ 2} \right)^{2}}
\label{tilde_g} \\
\tilde{s} \left( \Omega_{s} \right) &=& \sqrt{\pi} f_{s} \tau_{s} e^{- \left( \frac{\Omega_{s} \tau_{s}}{ 2} \right)^{2}}e^{-i\Omega_{s}\Delta t}
\label{tilde_s} 
\end{eqnarray}
Using eq.(\ref{lin_pol_burst_Gaussian}) in (\ref{c02_f}) the transition amplitude from the ground state to the second excited state for the burst signal becomes
 \begin{eqnarray}
C_{0\rightarrow 2}&=& - \frac{i}{2\pi\sqrt{24}}\left[ \int_{-\infty}^{+\infty} d\Omega \, \Omega \, \tilde{g}\left( \Omega \right) A_{\times} \int_{-\infty}^{+\infty} dt~~ e^{i(2\omega-\Omega)t}  + \int_{-\infty}^{+\infty} d\Omega_{s}\,\Omega_{s}\,  \tilde{s}\left( \Omega_{s} \right) A_s  \int_{-\infty}^{+\infty} dt~~ e^{i(2\omega-\Omega_{s})t} \right]\nonumber\\
&=& - \frac{i}{\sqrt{24}}\left[ \int_{-\infty}^{+\infty} d\Omega \, \Omega \, \tilde{g}\left( \Omega \right) A_{\times} \delta\left(2\omega-\Omega \right)  + \int_{-\infty}^{+\infty} d\Omega_{s}\,\Omega_{s}\,  \tilde{s}\left( \Omega_{s} \right) A_s  \delta \left(2\omega-\Omega_{s}\right) \right]
\label{gwbta1}
\end{eqnarray}
If $\Omega \neq \Omega_{s}$ only one of the delta functions can be picked up at a time by tunning the trapping frequency $\omega = 2 \pi \nu_{0}$ of the optical sensor in the LSD.  The corresponding probabilities for transition induced by the tensor and scalar modes are
\begin{eqnarray}
 P_{0\rightarrow 2} &=& \begin{cases} \frac{2\pi^{3}}{3}\left(A_{\times} f_{g} \right)^{2}\left(\frac{\nu_{0}}{\nu_{\rm max}}\right)^{2}  e^{-8 \pi^{2}\left(\frac{\nu_{0}}{\nu_{\rm max}}\right)^{2}},  \left[{\rm tensor \,\, mode}\right]\nonumber \\ 
    \frac{2\pi^{3}}{3}\left(A_{s} f_{s} \right)^{2}\left(\frac{\nu_{0}}{\nu_{\rm max}}\right)^{2}  e^{-8 \pi^{2}\left(\frac{\nu_{0}}{\nu_{\rm max}}\right)^{2}},  \left[{\rm scalar \,\, mode}\right]  
    \end{cases}\\
 \end{eqnarray}
where we have used (\ref{tilde_g},  \ref{tilde_s}) and assumed similar temporal duration of the burst $\tau_g \approx  \tau_s \sim \frac{1}{\nu_{\rm max}}$ for both modes.  The probabilities for the two modes differ only by the relative strengths of their signal amplitudes.  Both the transition probabilities are significant when the trapping frequency $\nu_{0}$ falls within the spread $\nu_{\rm max}$ determined by the temporal duration of the burst and reach a maximum at $(\frac{\nu_{0}}{\nu_{\rm max}}) = \frac{1}{2 \sqrt{2} \pi}$. \\

Similarly amplitude for transition form the first excited state to the third is
\begin{eqnarray}
C_{1\rightarrow 3}&=&   -i \left[ \frac{2 \sqrt{2} +1}{2 \sqrt{30}}\int_{-\infty}^{+\infty} d\Omega \, \Omega \, \tilde{g}\left( \Omega \right) A_{\times} \delta\left(2\omega-\Omega \right)  +  \frac{2 + \sqrt{3} }{2 \sqrt{30}}\int_{-\infty}^{+\infty} d\Omega_{s}\,\Omega_{s}\,  \tilde{s}\left( \Omega_{s} \right) A_s  \delta \left(2\omega-\Omega_{s}\right) \right]
\label{gwbta2}
\end{eqnarray}
resulting in the transition probabilities
\begin{eqnarray}
 P_{1\rightarrow 3} & = & \begin{cases} \frac{2\pi^{3}}{15}\left(3.828 \, A_{\times} f_{g}\right) ^{2}\left(\frac{\nu_{0}}{\nu_{\rm max}}\right)^{2}  e^{-8 \pi^{2}\left(\frac{\nu_{0}}{\nu_{\rm max}}\right)^{2}},  \left[{\rm tensor \,\, mode}\right]\nonumber \\ 
    \frac{2\pi^{3}}{15}\left(3.732 \, A_{s} f_{s} \right)^{2}\left(\frac{\nu_{0}}{\nu_{\rm max}}\right)^{2}  e^{-8 \pi^{2}\left(\frac{\nu_{0}}{\nu_{\rm max}}\right)^{2}},  \left[{\rm scalar \,\, mode}\right]  
    \end{cases}\\
 \end{eqnarray}
 which show almost the same characteristics as the $0 \to 2$ transition, but of course with reduced strengths.

A slightly more realistic representation of the burst signal can be realized if we modulate the gaussian wave-packets of the massless and massive modes with carrier frequencies satisfying the respective dispersion relations\footnote{We have used $\hbar = 1=c$ in this relation and,  as discussed earlier,  kept the frequencies for scalar and tensor modes distinct.}  $\Omega_{g0} = |\vec{k}|$ and $ \Omega_{s0} = \sqrt{m^{2}+k^{2}}$ as
\begin{eqnarray}
g \left(t\right) &=& f_{g} e^{- t^{2}/ \tau_{g}^{2}}  \,\sin \Omega_{g0}t  \nonumber\\
s \left(t\right) &=& f_{s} e^{- \left(t+\Delta t\right)^{2}/ \tau_{s}^{2}}  \, \sin \Omega_{s0}t
\label{burst_waveform_sine_Gaussian}
\end{eqnarray}
so that their Fourier transforms 
\begin{eqnarray}
\tilde{g} \left( \Omega \right) &=&  \int_{-\infty}^{+\infty} g(t) e^{i \Omega t} d \Omega 
= \frac{i \sqrt{\pi} f_{g} \tau_{g}}{2} \left[ e^{- \left(\Omega - \Omega_{g0}\right)^{2}\tau_{g}^{2}/4} - e^{- \left(\Omega + \Omega_{g0}\right)^{2}\tau_{g}^{2}/4} \right] 
\label{sine-Gaussian_Fourier_1}\\
\tilde{s} \left( \Omega_{s} \right) &=&   \int_{-\infty}^{+\infty} s(t) e^{i \Omega_{s} t} d \Omega_{s}=\frac{i \sqrt{\pi} f_{s} \tau_{s}}{2} \left[ e^{- i \left(\Omega_{s} - \Omega_{s0}\right) \Delta t}  \,\, e^{- \left(\Omega_{s} - \Omega_{s0}\right)^{2}\tau_{s}^{2}/4} -  e^{-i \left(\Omega_{s} + \Omega_{s0}\right) \Delta t} \,\, e^{- \left(\Omega_{s} + \Omega_{s0}\right)^{2}\tau_{s}^{2}/4}\right]\nonumber\\
\label{sine-Gaussian_Fourier_2}
\end{eqnarray}
have similar spread in the frequency domain $\frac{\Omega_{\rm max}}{2 \pi} = \nu_{\rm max} \sim \frac{1}{\tau_g} \approx  \frac{1}{\tau_s} $,  but are centered around the corresponding carrier frequencies.
Substituting these waveforms (\ref{sine-Gaussian_Fourier_1},\ref{sine-Gaussian_Fourier_2}) in equations (\ref{gwbta1}, \ref{gwbta2}),  we get the transition amplitudes  
\begin{eqnarray}
C_{0\rightarrow 2}\left[{\rm tensor\,mode}\right] &= &  \sqrt{\frac{ \pi}{6}}\left(\frac{A_{\times} f_{g}\, \omega\,  \tau_{g} }{2}\right)\left[ e^{- \left(2\omega - \Omega_{g0}\right)^{2}\tau_{g}^{2}/4} - e^{- \left(2\omega + \Omega_{g0}\right)^{2}\tau_{g}^{2}/4}\right] \\
C_{0\rightarrow 2} \left[{\rm scelar\, mode}\right] &=& \sqrt{\frac{ \pi}{6}}\left(\frac{A_{s} f_{s}\, \omega\,  \tau_{s} }{2}\right)\left[ e^{-i \left(2\omega - \Omega_{s0}\right)\Delta t} \,\,e^{- \left(2\omega - \Omega_{s0}\right)^{2}\tau_{s}^{2}/4} - e^{-i \left(2\omega + \Omega_{s0}\right)\Delta t} \,\,e^{- \left(2\omega + \Omega_{s0}\right)^{2}\tau_{s}^{2}/4}\right] \\
\label{gwptt1}
C_{1\rightarrow 3}\left[{\rm tensor\,mode}\right] &= & \left(2 \sqrt{2}+1\right) \sqrt{\frac{ \pi}{30}}\left(\frac{A_{\times} f_{g}\, \omega\,  \tau_{g} }{2}\right)\left[ e^{- \left(2\omega - \Omega_{g0}\right)^{2}\tau_{g}^{2}/4} - e^{- \left(2\omega + \Omega_{g0}\right)^{2}\tau_{g}^{2}/4}\right] \\
C_{1\rightarrow 3} \left[{\rm scelar\, mode}\right] &=& \left(2 + \sqrt{3}\right)\sqrt{\frac{ \pi}{30}}\left(\frac{A_{s} f_{s}\, \omega\,  \tau_{s} }{2}\right)\left[ e^{-i \left(2\omega - \Omega_{s0}\right)\Delta t} e^{- \left(2\omega - \Omega_{s0}\right)^{2}\tau_{s}^{2}/4} - e^{-i \left(2\omega + \Omega_{s0}\right)\Delta t} e^{- \left(2\omega + \Omega_{s0}\right)^{2}\tau_{s}^{2}/4}\right] \nonumber\\
\label{gwptt2}
\end{eqnarray}
When the trapping frequency $\omega = 2 \pi \nu_{0}$ satisfies the condition $\nu_{0} = \frac{\nu_{g0}}{2} = \frac{\Omega_{g0}}{4\pi}$ with the carrier frequency $\Omega_{g0}$ of the tensor mode or the condition $\nu_{0} = \frac{\nu_{s0}}{2} = \frac{\Omega_{s0}}{4\pi}$ with the carrier frequency $\Omega_{s0}$ of the scelar mode,  the corresponding transition probabilities are
\begin{eqnarray}
P_{0\rightarrow 2} &=& \begin{cases} \frac{\pi}{96}\left(A_{\times} f_{g} \right)^{2}\chi_{g}^{2}  \left[1 + e^{-2 \chi_{g}^{2}} - 2 e^{- \chi_{g}^{2}}\right],  \left[{\rm tensor \,\, mode}\right]\nonumber \\ 
    \frac{\pi}{96}\left(A_{s} f_{s} \right)^{2}\chi_{s}^{2}  \left[1 + e^{-2 \chi_{s}^{2}} - 2 \cos\left(2 \chi_{s} \frac{\Delta t}{\tau_{s}}\right)e^{- \chi_{s}^{2}}\right]  \left[{\rm scalar \,\, mode}\right]  
    \end{cases}\\
    \label{TP_SG_02}
\end{eqnarray}
\begin{eqnarray}
P_{1\rightarrow 3} &=& \begin{cases} \frac{\pi}{480}\left(3.828 \, A_{\times} f_{g} \right)^{2}\chi_{g}^{2}  \left[1 + e^{-2 \chi_{g}^{2}} - 2 e^{- \chi_{g}^{2}}\right],  \left[{\rm tensor \,\, mode}\right]\nonumber \\ 
    \frac{\pi}{480}\left(3.732 \, A_{s} f_{s} \right)^{2}\chi_{s}^{2}  \left[1 + e^{-2 \chi_{s}^{2}} - 2 \cos\left(2 \chi_{s} \frac{\Delta t}{\tau_{s}}\right)e^{- \chi_{s}^{2}}\right]  \left[{\rm scalar \,\, mode}\right]  
    \end{cases}\\
 \label{TP_SG_13}
\end{eqnarray}
where $\chi_{g} = \frac{2 \pi \nu_{g0}}{\nu_{\rm max}} $ and $\chi_{s} = \frac{2 \pi \nu_{s0}}{\nu_{\rm max}}$.  The factors in the square brackets in (\ref{TP_SG_02}, \ref{TP_SG_13}) quickly grow from zero to their maximum value ``one'' as $\chi_{g}$ and $\chi_{s}$ increase from zero.  The cosine term appearing for the longitudinal scalar mode owing to their relative delay $\Delta t$ in arrival has no significant effect in the transition probabilities.  Thus the overall behaviour of all these transition probabilities are primarily determined by the $\chi_{g}^{2}$ or $\chi_{s}^{2}$ factor in front,  making them increasingly significant at higher frequency values,  similar to the case of periodic signals.   
Since at higher trapping frequencies the efficiency of the levitated sensor detector gets better they may prove to be an apt tool to detect such transitions induced by GW bursts.  Comparing the probabilities for a given transition induced by the tensor and scalar modes one can draw similar inferences about the relative strength of the two modes in case of a brust signal as we have discussed  in the previous section in case of a periodic signal.
\section{Discussion}
In this paper we have considered the response of the recently proposed levitated sensor detector(LSD) of gravitational wave (GW) to an incoming signal in the framework of modified theories of gravity(MTG).  In addition to the massless tensor polarization modes of GW present in standard general relativity (GR) the MTG predicts a massive longitudinal scalar polarization mode.  Being massive it has a delay in arrival time or phase lag compard to massless tensor modes.  Being logitudinal it is also expected to move a test mass out of the transverse plane in which the effects of the tensor modes are confined.  These distinctive features of the scalar mode,  if identified in GW data,  can provide observational evidance in favour of MTG over standard GR,  thereby changing our current understanding of the innate nature of gravity itself.  Since a stand-alone GW detector can not resolve the propagation direction of an incoming signal to extract polarization information therein,  a network of widely seperated and simultaneously operating GW detectors is necessary.  

The currently operational interferometric detector like Advanced LIGO and VIRGO have been successfully detecting transient signals for nearly a decade and some more are currently under way.  However owing to their large-scale design and huge budget,  developing and managing a fully operational interferometric detector is a challenging task and only a handful of such sites can be built.  In contrast the levitated sensor detector(LSD) works on the principle of GW indeced resonant transition in a quantum harmonic oscillator,  realized by optically trapping a dielectric nanosphere in the harmonic potential at an antinode of a optical cavity.  Unlike the traditional resonant bar detectors,  the frequency of the trapping potential in LSD is variable,  allowing for a scanning stratagy to pick up the resonant frequency.  Moreover its compact design makes multiple detector sites as well as several detector orientations at a given site entirely plausible.  Therefore anticipating that LSDs may prove to be ideal for picking up any possible signature of massive longitudinal scalar mode in a GW signal we have constructed the quantum mechanics of its response to GW signals in an generic MTG framework in this paper.  

To incorporate the effect of the longitudinal scalar mode we have considered the dynamics of the sensor mass of the LSD in three-dimensions.  The rest frame of the LSD is an Earth-bound proper detector frame.  Since in a LSD set up along with the sensor mass also the antinode position gets affected by the passing GW we analysed the geometry of its optical cavity to demonstrate that the motion of the sensor mass is indeed governed by a geodesic deviation equation at a classical level in this proper detector frame.  In the long wavelength and low velocity limit this geodesic deviation equation takes the form of a Newtonian force equation for the sensor mass.  We quantiz the corresponding Hamiltonian to obtain a quantum mechanical description of the system.  Apart from the formal solution depicting the quantum dynamics of the sensor mass perturbed by scalar and tensor modes we have computed the probabbilities for its first two most prominant resonant transitions induced by GW.
 Our results show that the probability of a given transition depends primarily on the amplitudes of the polarization modes and their frequencies.  It increases quadratically with the frequency of a periodic signal and also with the carrier frequency of a burst signal.  This is encouraging since LSD claims to have improved sensitivity at the higher frequency side.

Once a network of LSDs becomes operational their combined data should be able to sufficiently resolve the source direction so that the transverse plane can be identified.  With multiple orientations of such LSDs at a given site,  one should be able to filter out the effect of a possible massive scalar polarization mode where their delayed arrivel time or phase lag compared to the tensor modes will act as an identifier.  From this data the mass scale associated with the scalar mode characterizing how modified a specific MTG is from the standard GR can be estimated.  For GW signal coming from a specific source if both modes arrive with the same frequency from the ratio of their corresponding transition probabilities one can obtain the ratio of their amplitudes which should be consistent for $P_{0 \to 2}$ and $P_{1 \to 3}$ transitions.  This will be useful in modeling the GW source. 
On the other hand if the tensor and scalar modes differ in frequency and both the resonant transitions can be picked up by a properly coordinated scanning stratagy at several detector sites the difference among the resonance points can act as an additional identifier for the scalar mode.  It will also help estimate the mass scale of the scalar mode that can be corroborated against the estimation made from phase lag/ arrival time delay mentioned earlier. The scope to probe all this new physics in GW data depends on the wide tunability of the LSD,  its compact design and economic viability making it a very important alternative to the currently operating ground-based interferometric GW detectors.

\begin{appendix}
\section{Formal Solution}\label{AppA}
In this appendix we shall present the formal solution which depicts the dynamics of the sensor mass in the levitated sensor detector (LSD) interacting with the massless tensor modes and massive longitudinal scalar mode of gravitational wave (GW) in the modified theory of gravity (MTG) framework.  In terms of the lowering and raising operators $a_{i}(t)$ and $a^{\dagger}_{i}(t)$ defined in (\ref{operator}) and satisfying the algebra
\begin{eqnarray} \label{operator1}
\left[a_{j}(t),  a^{\dagger}_{k}(t)\right] = \delta_{jk};  \,\, \,\,\,
\left[a_{j}(t),a_{k}(t)\right] = 0=
\left[a^{\dagger}_{j}(t),a^{\dagger}_{k}(t)\right] 
\end{eqnarray} 
the quantum mechanical Hamiltonian (\ref{hamilton_HO}) describes the system and dictates its time evolution through the Heisenberg equations (\ref{operator3}) for the lowering operators $a_{i}(t)$ which, upon using the algebra (\ref{operator1}),  becomes
\begin{eqnarray}\label{operator3a}
\frac{da_{i}(t)}{dt} = -\frac{i\omega}{2}\left(a_{i}-a^{\dagger}_{i}\right)+\frac{\dot{h}_{ij}}{2}a_{j} 
\end{eqnarray}
Its complex conjugate (C.C) gives the time evolution for the raising operators $a^{\dagger}_{i}(t)$.  
These operators can be expressed in terms of their initial values $a_{j}(t =0) = a_{j}(0)$ etc,  using the Bogoliubov transformations
\begin{eqnarray} \label{operator2}
a_{j}(t)&=& \mathcal{U}_{jk}(t)a_{k}(0)+\mathcal{V}_{jk}(t)a^{\dagger}_{k}(0) \nonumber\\
a^{\dagger}_{j}(t) &=& a^{\dagger}_{k}(0)\bar{\mathcal{U}}_{kj}(t)+a_{k}(0)\bar{\mathcal{V}}_{kj}(t)
\end{eqnarray}
such that the time dependence is accessed through the $3\times 3$ Bogoliubov coefficients $\mathcal{U}_{jk}(t)$ and $\mathcal{V}_{jk}(t)$. Their initial values are $\left(\mathcal{U}_{jk}(0)=\delta_{jk}\,, \mathcal{V}_{jk}(0)=0\right)$ and they satisfy the algebra 
$\mathcal{U}\mathcal{V}^{T}=\mathcal{U}^{T}\mathcal{V};\,\, \mathcal{U}\mathcal{U}^{\dagger}-\mathcal{V}\mathcal{V}^{\dagger}=I,\,$
because of eq.(\ref{operator1}).  Here $I$ is the $3 \times 3$ identity matrix,  $T$ and $\dagger$ denote transpose and conjugate transpose respectively.  

Utilizing the Bogoliubov coefficients ($\mathcal{U},\mathcal{V}$) we define a new pair of variables ($\zeta,\eta$) 
\begin{align}\label{bogo2}
\zeta_{jk}=\mathcal{U}_{jk}-\mathcal{V}^{\dagger}_{jk};~~~\eta_{jk}=\mathcal{U}_{jk}+\mathcal{V}^{\dagger}_{jk}
\end{align} 
to express the dynamical evolution of our system as
\begin{eqnarray}
\label{zeta_t}
\frac{d\zeta_{jk}}{dt}&=& -i\omega\eta_{jk}  -\frac{1}{2}\dot{h}_{jp}\zeta_{pk}\\
\label{xi_t}
\frac{d\eta_{jk}}{dt}&=&-i\omega\zeta_{jk}+\frac{1}{2}\dot{h}_{jp}\eta_{pk} ~~~~.
\end{eqnarray} 
where equations (\ref{operator2}) and (\ref{bogo2}) have been used in equation (\ref{operator3a}) and its   C.C.   

We expand $ \zeta_{jk}$ and $\eta_{jk} $ using a suitable basis $\left\{\wp_{jk}^{M}\,,\,\, \left(M = 1,\,2,\, ...,\, 9\right)\right\}$ spanning the space of $3 \times 3$ matrices as
\begin{eqnarray}
\zeta_{jk}(t)&=&\sum ^{9}_{M=1}\mathcal{K}_{M}\left(t\right)\wp^{M}_{jk}\nonumber\\
\eta_{jk}(t)&=&\sum ^{9}_{M=1}\mathcal{Q}_{M}\left(t\right)\wp^{M}_{jk}  \label{zeta}
\end{eqnarray}
where $\left\{\wp_{jk}^{M}\,,\,\, \left(M = 1,\,2,\, ...,\, 9\right)\right\}$ has,  along with the three polarization matrices $\left( \wp^{1}_{jk},\, \wp^{2}_{jk}, \, \wp^{3}_{jk}\right)$ given in (\ref{pol_basis}),  six other independent matrices 
\begin{align*}
\wp^{4}_{jk}=
\begin{pmatrix}
0 & 0 & 1 \\
0 & 0 & 0 \\
1 & 0 & 0
\end{pmatrix}, \,\, 
\wp^{5}_{jk}=
\begin{pmatrix}
0 & 0 & 0 \\
0 & 0 & 1 \\
0 & 1 & 0
\end{pmatrix},\,\, 
\wp^{6}_{jk}=
\begin{pmatrix}
1 & 0 & 0 \\
0 & 1 & 0 \\
0 & 0 & 0
\end{pmatrix} , 
\end{align*}
\begin{align}\label{pol_basis_extra}
\wp^{7}_{jk}=
\begin{pmatrix}
~0 & 1 & 0 \\
-1 & 0 & 0 \\
~0 & 0 & 0
\end{pmatrix},\,\,
\wp^{8}_{jk}=
\begin{pmatrix}
~0 & 0 & 1 \\
~0 & 0 & 0 \\
-1 & 0 & 0
\end{pmatrix},
\wp^{9}_{jk}=
\begin{pmatrix}
0 & ~0 & 0 \\
0 & ~0 & 1 \\
0 & -1 & 0
\end{pmatrix}.
\end{align}
Using the generic form of the GW signal 
$h_{jk}\left(t\right) = \sum_{I} h_{I}\left(t\right) \wp^{I}_{jk},\,\, \left(I = 1,\,2,\,3\right)$
and the expansion (\ref{zeta}) the coupled differential equations (\ref{zeta_t}, \ref{xi_t}) for the $3 \times 3$ matrices reduce to a set of first-order differential equations for the coefficients $\mathcal{K}_{M}$ and $\mathcal{Q}_{M}$:
\begin{eqnarray*}
\dot{\mathcal{K}_{1}}&=&-i\omega \mathcal{Q}_{1} - \frac{1}{2}\left(\mathcal{K}_{6}\dot{h}_{\times}+\mathcal{K}_{7}\dot{h}_{+}\right)\nonumber\\
\dot{\mathcal{K}_{2}}&=&-i\omega \mathcal{Q}_{2}- \frac{1}{2}\left(\mathcal{K}_{7}\dot{h}_{\times}-\mathcal{K}_{6}\dot{h}_{+}\right)\nonumber\\
\dot{\mathcal{K}_{3}}&=&-i\omega \mathcal{Q}_{3}-\frac{\mathcal{G}_{3}\dot{h}_{s}}{\sqrt{2}}\nonumber\\
\dot{\mathcal{K}_{4}}&=&-i\omega \mathcal{Q}_{4}-\frac{\dot{h}_{\times}}{4}\left(\mathcal{K}_{5}+\mathcal{K}_{9}\right) -\frac{\dot{h}_{+}}{4}\left(\mathcal{K}_{4}+\mathcal{K}_{8}\right)-\frac{\dot{h}_{s}}{2\sqrt{2}}\left(A_{4}-A_{8}\right)\nonumber\\
\dot{\mathcal{K}_{5}}&=&-i\omega \mathcal{Q}_{5}-\frac{\dot{h}_{\times}}{4}\left(\mathcal{K}_{4}+\mathcal{K}_{8}\right)+\frac{\dot{h}_{+}}{4}\left(\mathcal{K}_{5}+\mathcal{K}_{9}\right)-\frac{\dot{h}_{s}}{2\sqrt{2}}\left(\mathcal{K}_{5}-\mathcal{K}_{9}\right)\nonumber\\
\dot{\mathcal{K}_{6}}&=&-i\omega \mathcal{Q}_{6}-\frac{1}{2}\left(\mathcal{K}_{1}\dot{h}_{\times}+\mathcal{K}_{2}\dot{h}_{+}\right)\nonumber\\
\dot{\mathcal{K}_{7}}&=&-i\omega \mathcal{Q}_{7}\frac{1}{2}\left(\mathcal{K}_{2}\dot{h}_{\times}-\mathcal{K}_{1}\dot{h}_{+}\right)\nonumber\\
\dot{\mathcal{K}_{8}}&=&-i\omega \mathcal{Q}_{8}-\frac{\dot{h}_{\times}}{4}\left(\mathcal{K}_{5}+\mathcal{K}_{9}\right)-\frac{\dot{h}_{+}}{4}\left(\mathcal{K}_{4}+\mathcal{K}_{8}\right)+\frac{\dot{h}_{s}}{2\sqrt{2}}\left(\mathcal{K}_{4}-\mathcal{K}_{8}\right)\nonumber\\
\dot{\mathcal{K}_{9}}&=&-i\omega \mathcal{Q}_{9}-\frac{\dot{h}_{\times}}{4}\left(\mathcal{K}_{4}+\mathcal{K}_{8}\right) +\frac{\dot{h}_{+}}{4}\left(\mathcal{K}_{5}+\mathcal{K}_{9}\right)+\frac{\dot{h}_{s}}{2\sqrt{2}}\left(\mathcal{K}_{5}-\mathcal{K}_{9}\right)
\end{eqnarray*}
\begin{eqnarray}\label{amplitudes}
\dot{\mathcal{Q}_{1}}&=&-i\omega \mathcal{K}_{1}-\frac{1}{2}\left(\mathcal{Q}_{6}\dot{h}_{\times}+\mathcal{Q}_{7}\dot{h}_{+}\right)\nonumber\\
\dot{\mathcal{Q}_{2}}&=&-i\omega \mathcal{K}_{2}+\frac{1}{2}\left(\mathcal{Q}_{7}\dot{h}_{\times}-\mathcal{Q}_{6}\dot{h}_{+}\right)\nonumber\\
\dot{\mathcal{Q}_{3}}&=&-i\omega \mathcal{K}_{3}-\frac{\mathcal{Q}_{3}\dot{h}_{s}}{\sqrt{2}}\nonumber\\
\dot{\mathcal{Q}_{4}}&=&-i\omega \mathcal{K}_{4} -\frac{\dot{h}_{\times}}{4}\left(\mathcal{Q}_{5}+\mathcal{Q}_{9}\right) -\frac{\dot{h}_{+}}{4}\left(\mathcal{Q}_{4} +\mathcal{Q}_{8}\right)-\frac{\dot{h}_{s}}{2\sqrt{2}}\left(\mathcal{Q}_{4}-\mathcal{Q}_{8}\right)\nonumber\\
\dot{\mathcal{Q}_{5}}&=&-i\omega \mathcal{K}_{5}-\frac{\dot{h}_{\times}}{4}\left(\mathcal{Q}_{4}+\mathcal{Q}_{8}\right)+  \frac{\dot{h}_{+}}{4}\left(\mathcal{Q}_{5}+\mathcal{Q}_{9}\right)-\frac{\dot{h}_{s}}{2\sqrt{2}}\left(\mathcal{Q}_{5}-\mathcal{Q}_{9}\right)\nonumber\\
\dot{\mathcal{Q}_{6}}&=&-i\omega \mathcal{K}_{6} -\frac{1}{2}\left(\mathcal{Q}_{1}\dot{h}_{\times}+\mathcal{Q}_{2}\dot{h}_{+}\right)\nonumber\\
\dot{\mathcal{Q}_{7}}&=&-i\omega \mathcal{K}_{7}+\frac{1}{2}\left(\mathcal{Q}_{2}\dot{h}_{\times}-\mathcal{Q}_{1}\dot{h}_{+}\right)\nonumber\\
\dot{\mathcal{Q}_{8}}&=&-i\omega \mathcal{K}_{8}-\frac{\dot{h}_{\times}}{4}\left(\mathcal{Q}_{5}+\mathcal{Q}_{9}\right)-\frac{\dot{h}_{+}}{4}\left(\mathcal{Q}_{4}+\mathcal{Q}_{8}\right)+\frac{\dot{h}_{s}}{2\sqrt{2}}\left(\mathcal{Q}_{4}-\mathcal{Q}_{8}\right)\nonumber\\
\dot{\mathcal{Q}_{9}}&=&-i\omega \mathcal{K}_{9}-\frac{\dot{h}_{\times}}{4}\left(\mathcal{Q}_{4}+\mathcal{Q}_{8}\right) +\frac{\dot{h}_{+}}{4}\left(\mathcal{Q}_{5}+\mathcal{Q}_{9}\right)+\frac{\dot{h}_{s}}{2\sqrt{2}}\left(\mathcal{Q}_{5}-\mathcal{Q}_{9}\right)
\end{eqnarray}
Since $|h_{I}\left(t \right)|\ll 1$,  this set of equations (\ref{amplitudes}) can be solved iteratively around their $h_{I}(t)=0$ solutions.  As an appropriate boundary condition on $h_{I}\left(t \right)$,  we assume that the GW hits the system at $t = 0$,  so that
$h_{I}\left(t \right)=0,\, {\rm for}\,t\leq 0$.
The first-order solutions
\begin{align}
\mathcal{K}_{1}=\mathcal{Q}_{1}= -\frac{1}{2}\int^{t}_{0}\dot{h}_{\times}e^{-i\omega t}dt,~~~~
\mathcal{K}_{2}=\mathcal{Q}_{2}= -\frac{1}{2}\int^{t}_{0}\dot{h}_{+}e^{-i\omega t}dt,~~~~
\mathcal{K}_{3}=\mathcal{Q}_{3}= \frac{1}{\sqrt{2}}e^{-i\omega t}-\frac{1}{2}\int^{t}_{0}\dot{h}_{s}e^{-i\omega t}dt \nonumber \\
\mathcal{K}_{4} = \mathcal{Q}_{4}= 0 ;~~
\mathcal{K}_{5} = \mathcal{Q}_{5}= 0;~~
\mathcal{K}_{6} = \mathcal{Q}_{6}= e^{-i\omega t};~~
\mathcal{K}_{7} = \mathcal{Q}_{7}= 0;~~
\mathcal{K}_{8} = \mathcal{Q}_{8}= 0;~~
\mathcal{K}_{9} = \mathcal{Q}_{9}= 0
\label{zero_order}
\end{align}
are used in (\ref{zeta}) to obtain the matrix pair ($\zeta_{jk},\eta_{jk}$) which are then further substituted in (\ref{operator2}) via (\ref{bogo2}) to obtain the time-evolution of the raising and lowering operators $a_{j}(t)$ and $a^{\dagger}_{j}(t)$ in terms of their initial values, $a_{j} (0)$ and $a^{\dagger}_{j}(0)$.  
Also using (\ref{operator}) the initial position and momentum expectation values $ \langle\hat{\xi}_{i}\rangle_{t=0}=(\xi_{10},\xi_{20},\xi_{30})$
and $\langle\hat{p}_{i}\rangle_{t=0}=(P_{10},P_{20},P_{30})$ can be expressed in terms of ($a_{j}(0) ~,~a^{\dagger}_{j}(0)$). Therefore we can finally express the time evolution of the position expectation values $<\hat{\xi}_{i}(t)>$ in terms of their initial values  
\begin{eqnarray}
\langle \hat{\xi}_{1}(t)\rangle&=& \xi_{10}\cos\omega t+ \frac{P_{10}}{m_{s}\omega}\sin\omega t -\frac{\xi_{10}}{2}\int^{t}_{0}\dot{h}_{+}\cos\omega t' dt'+\frac{P_{10}}{2m_{s}\omega}\int^{t}_{0}\dot{h}_{+}\sin\omega t' dt'-\frac{\xi_{20}}{2}\int^{t}_{0}\dot{h}_{\times}\cos\omega t' dt'\nonumber\\
                & &+ \frac{P_{20}}{2m_{s}\omega}\int^{t}_{0}\dot{h}_{\times}\sin\omega t' dt'\nonumber\\
\langle \hat{\xi}_{2}(t)\rangle&=& \xi_{20}\cos\omega t+ \frac{P_{20}}{m_{s}\omega}\sin\omega t -\frac{\xi_{10}}{2}\int^{t}_{0}\dot{h}_{\times}\cos\omega t' dt' -\frac{P_{10}}{2m_{s}\omega}\int^{t}_{0}\dot{h}_{\times}\sin\omega t' dt'-\frac{\xi_{20}}{2}\int^{t}_{0}\dot{h}_{+}\cos\omega t' dt' \nonumber\\
                & &- \frac{P_{20}}{2m_{s}\omega}\int^{t}_{0}\dot{h}_{+}\sin\omega t' dt'\nonumber\\
\langle \hat{\xi}_{3}(t)\rangle&=& \xi_{30}\cos\omega t + \frac{P_{30}}{m_{s}\omega}\sin\omega t -\frac{\xi_{30}}{\sqrt{2}}\int^{t}_{0}\dot{h}_{s}\cos\omega t' dt' -\frac{P_{30}}{\sqrt{2}m_{s}\omega}\int^{t}_{0}\dot{h}_{s}\sin\omega t' dt' 
\label{expectation_values}
\end{eqnarray}
The equations in (\ref{expectation_values}) are the formal solution describing the dynamics of the nanosphere sensor in the LSD exposed to incoming GW signals,  in terms of its position expectation values.  

Since the sensor mass $m_{{\rm s}}$ is trapped in a harmonic potential in the optical cavity the underlying system is a quantum harmonic oscillator forced by the GW signal.  Therefore these expectation values mimic the behaviour of a classical forced harmonic oscillator as per the Ehrenfest theorem.  As anticipated the effect of the longitudinal scalar mode $h_{s}\left(t \right)$ is limited to the GW signal propagation direction $z$ and the usual tensor modes $h_{\times}\left(t \right)$ and $h_{+}\left(t \right)$ affect the oscillator only in the transverse plane.  The limiting scenario of no incoming GW signals can be obtained by implimenting the condition,  $h_{I}\left(t\right) = 0, \,\left(I = \times,  +,  {\rm s}\right)$ for all time $t$ in (\ref{expectation_values}),  which yields the standard three-dimensional harmonic oscillator solution
\begin{eqnarray}
\langle \hat{\xi}_{1}(t)\rangle = \xi_{10}\cos\omega t+ \frac{P_{10}}{m_{s}\omega}\sin\omega t, \,\,\,\,
\langle\hat{\xi}_{2}(t)\rangle= \xi_{20}\cos\omega t+ \frac{P_{20}}{m_{s}\omega}\sin\omega t, \,\,\,\,
\langle\hat{\xi}_{3}(t)\rangle= \xi_{30}\cos\omega t + \frac{P_{30}}{m_{s}\omega}\sin\omega t 
\end{eqnarray}
providing a basic consistency check for (\ref{expectation_values}).

\end{appendix}
\section*{Acknowledgment} \noindent 
RD thanks the Council of Scientific and Industrial Research (CSIR),  Govt. of India,  for financial support. The authors would like to acknowledge the hospitality of IUCAA,  Pune where part of the work has been carried out.  AS also thanks his collegues Dr.  Arunabha Adhikari and Dr.  Bibhas Bhattacharyya for valuable input and suggestions.

\end{document}